\documentclass[twocolumn]{jpsj3} %% two-column layout
\usepackage{graphicx}
\usepackage{bm}

\title{Emergence of Novel Antiferromagnetic Order Intervening between Two Superconducting Phases in LaFe(As$_{1-x}$P$_x$)O: $^{31}$P-NMR Studies }

\author{
Hidekazu \textsc{Mukuda}$^{1}$\thanks{E-mail: mukuda@mp.es.osaka-u.ac.jp}, Fuko \textsc{Engetsu}$^{1}$, Takayoshi \textsc{Shiota}$^{1}$,  Kwing To \textsc{Lai}$^{2}$,\\ Mitsuharu \textsc{Yashima}$^{1,4}$, Yoshio \textsc{Kitaoka}$^{1}$, Shigeki \textsc{Miyasaka}$^{2}$\thanks{E-mail: miyasaka@phys.sci.osaka-u.ac.jp}, and Setsuko \textsc{Tajima}$^{2}$
}

\inst{
$^{1}$Graduate School of Engineering Science, Osaka University, Toyonaka, Osaka 560-8531, Japan \\
$^{2}$Graduate School of Science, Osaka University, Toyonaka, Osaka 560-0043, Japan \\
}

\abst{
Systematic $^{31}$P-NMR studies of LaFe(As$_{1-x}$P$_x$)O compounds have revealed the emergence of a novel antiferromagnetic ordered phase (AFM-2) at 0.4$\le x \le$0.7 that intervenes between two superconductivity (SC) phases. This AFM-2 phase with N\'eel temperature $T_{\rm N}$= 35 K for $x$=0.6 is in strong contrast to the AFM order (AFM-1) at $x$=0 exhibiting $T_{\rm N}$ of 140 K. Previous $^{31}$P-NMR studies of LaFe(As$_{1-x}$P$_x$)(O$_{1-y}$F$_{y}$) have revealed that $T_{\rm c}$ reaches a maximum of 24 K for $x$=0.6 as a result of the marked enhancement of AFM spin fluctuations at low energies due to electron doping by the flourine substitution of $y$=0.05 for oxygen. The reason for this unexpected result has been found in the present work,  that is,  the emergence of AFM-2 at 0.4 $\le x \le$0.7 without electron doping. We note that AFM spin fluctuations arising from interband nesting on the $d_{XZ}$/$d_{YZ}$ orbits must be a key factor for the occurrence of SC around AFM-2. 
}

\recdate{\today}
\newpage
\begin{document}
\maketitle

%\pacs{74.70.Xa, 74.25.Ha, 76.60.-k}

%\section{Introduction}

Iron (Fe) oxypnictide LaFeAsO with an orthorhombic structure exhibits antiferromagnetic (AFM) order, and the substitution of F$^-$ for O$^{2-}$ induces superconductivity (SC) with a maximum transition temperature of $T_{\rm c}$=26 K in LaFeAsO$_{1-y}$F$_y$~\cite{Kamihara2008}. The isostructural compound LaFeP(O$_{1-y}$F$_{y}$) with P substituted for As also reveals the SC transition at $T_{\rm c}$=4 - 7 K, which is lower than that in the case of LaFeAs(O,F)~\cite{Kamihara2006}.  In Fe-pnictide superconductors, $T_{c}$ reaches a maximum of 55 K~\cite{Ren1,Ren2} when a FeAs$_{4}$ block forms a nearly regular tetrahedral structure~\cite{C.H.Lee}: The optimal values of the As-Fe-As bonding angle ($\alpha$), the height of pnictogen ($h_{\rm Pn}$) from the Fe plane, and the $a$-axis length ($a$) are 109.5$^\circ$~\cite{C.H.Lee}, $\sim$1.38 \AA~\cite{Mizuguchi}, and $\sim$3.9 \AA~\cite{Ren2}, respectively. In this context, since the substitution of P for As  makes the $a$-axis length smaller, $\alpha$ wider, and $h_{\rm Pn}$ smaller than the optimal values for high-$T_{\rm c}$ Fe pnictides, it is anticipated that $T_{\rm c}$ might decrease monotonically as $x$ increases in solid solution compounds LaFe(As$_{1-x}$P$_x$)(O$_{1-y}$F$_{y}$). 
Unexpectedly, $T_{\rm c}$ exhibits a nonmonotonic variation with $x$ in LaFe(As$_{1-x}$P$_x$)(O$_{1-y}$F$_{y}$) compounds\cite{Saijo,Miyasaka,Lai}. 
Previous $^{31}$P-NMR studies of these compounds have revealed that $T_{\rm c}$ reaches its respective maxima of 27 and 24 K for $x$=0.4 with $y$=0.1 and for $x$=0.6 with $y$=0.05, as a result of the marked enhancement of AFM spin fluctuations(AFMSFs) at low energies~\cite{Mukuda_PRB2014}.
The result provides clear evidence that $T_{\rm c}$ is enhanced by AFMSFs at low energies even though the lattice parameters deviate from their optimum values. 
However, another question should be addressed: Why are AFMSFs enhanced despite the fact that the lattice parameters of the compounds are far from those of the AFM mother compound LaFeAsO. 

In this Letter, we report the results of our $^{31}$P-NMR studies that a novel AFM ordered phase (AFM-2) emerges at  0.4 $\le x \le$0.7, intervening between two SC phases (SC-1 and SC-2) in LaFe(As$_{1-x}$P$_x$)O.  
The $^{31}$P-NMR Knight shift indicates the appearance of a sharp density of states (DOS) at the Fermi level derived from  a $d_{3Z^2-r^2}$ orbit, which is less relevant with the onset of SC-2. 
On the other hand,  AFMSFs arising from interband nesting on $d_{XZ}$/$d_{YZ}$ orbits are mainly responsible for the occurrence of SC around AFM-2. 

%\section{Experimental Procedures}

Polycrystalline samples of LaFe(As$_{1-x}$P$_x$)O were synthesized by the solid-state reaction method~\cite{Saijo,Miyasaka,Lai,Lai_submitted}. Powder X-ray diffraction measurements indicated that the lattice parameters of LaFe(As$_{1-x}$P$_x$)O exhibit a monotonic variation with $x$~\cite{Lai_submitted}. 
$^{31}$P-NMR($I$=1/2) measurement was performed on coarse powder samples of LaFe(As$_{1-x}$P$_x$)O with nominal contents $x$=0.3, 0.4, 0.5, 0.6, 0.7, 0.8, and 1.0.  
The $^{31}$P-NMR spectra in the AFM ordered state were obtained by sweeping a magnetic field  at a fixed frequency $f_0$=107 MHz. 
The Knight shift $K$ in the normal state was measured at a magnetic field of $\sim$11.95 T, which was calibrated using a resonance field of $^{31}$P in H$_3$PO$_4$. 
The nuclear spin lattice relaxation rate $(1/T_1)$ of $^{31}$P-NMR was obtained at a field of $\sim$11.95 T by fitting a recovery curve for $^{31}$P nuclear magnetization to a single exponential function, $m(t)\equiv (M_0-M(t))/M_0=\exp \left(-t/T_1\right)$. 
Here, $M_0$ and $M(t)$ are the nuclear magnetizations for a thermal equilibrium condition and at time $t$ after a saturation pulse, respectively. 

%fig1------------------------------------------------
\begin{figure}[htbp]
\centering
\includegraphics[width=8cm]{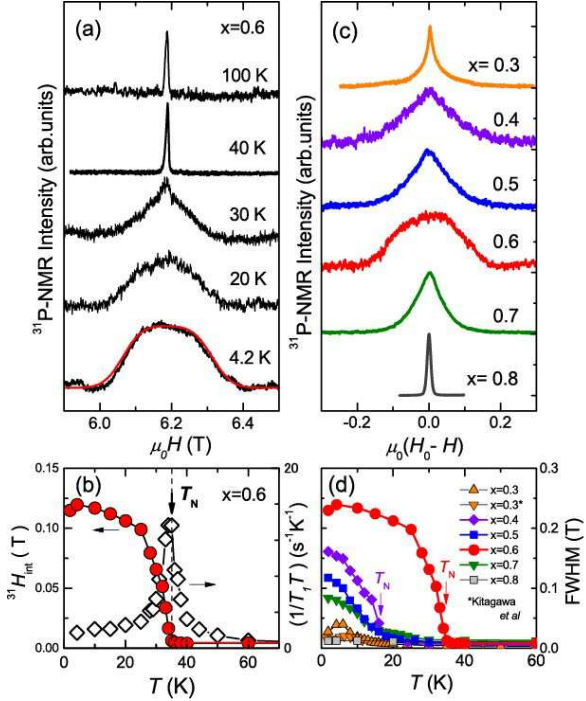}
\caption[]{(Color online) 
(a)  $T$ dependence of $^{31}$P-NMR spectra at $x$=0.6 of LaFe(As$_{1-x}$P$_x$)O.
The rectangle-like spectral shape indicates a commensurate AFM order. 
Solid curve is the simulation at $T$=4.2 K.  
%The rectangle-like spectral shape is characteristic of a randomly oriented powder where a commensurate AFM order takes place, as shown by the simulation at $T$=4.2 K.  
(b) $T$ dependences of $^{31}H_{\rm int}$ that is proportional to $M_{\rm AFM}$ and of $(1/T_1T)$ at $x$=0.6. Both results point to the emergence of AFM-2 with $T_{\rm N}$= 35 K.  (c) $x$ dependence of $^{31}$P-NMR spectra at $T$=1.9 K  and (d)  $T$ dependences of full-width at half maximum (FWHM) for each $x$, which provide evidence of AFM-2 occurring at $0.4\le x \le0.7$ but not at $x$=0.3 or 0.8. 
} 
\label{NMR_spectra}
\end{figure}
%------------------------------------------------

Figure~\ref{NMR_spectra}(a) shows the temperature ($T$) dependence of the $^{31}$P-NMR spectrum at $x$=0.6, which exhibits significant broadening below 35 K. At 4.2 K, the $^{31}$P-NMR spectrum indicates a rectangle-like spectral shape, which is characteristic of a randomly oriented powder where a commensurate AFM order takes place~\cite{Kinouchi_PRB_2013}. The $^{31}$P nucleus experiences a uniform off-diagonal internal hyperfine field, $^{31}\!H_{\rm int}$, associated with a stripe-type AFM order of Fe-$3d$ spins~\cite{Kitagawa1}. 
Actually, the spectrum can be simulated by assuming $^{31}\!H_{\rm int}\simeq$0.12$\pm$0.05 T at 4.2 K, as shown by the solid curve in Fig. \ref{NMR_spectra}(a).  
By using the relation  $^{i}\!H_{\rm int}$=$^{i}\!A_{\rm hf}M_{\rm AFM}$, an AFM moment $M_{\rm AFM}$ at the Fe site is estimated to be $\sim0.18(\pm0.07)\mu_{\rm B}$ by assuming the hyperfine-coupling constant  at the $^{75}$As site, $^{75}\!A_{\rm hf}$= 2.0$\sim$2.5 T/$\mu_{\rm B}$ in LaFeAsO~\cite{hf_LaFeAsO}, and the ratio $^{75}\!A_{\rm hf}/^{31}\!A_{\rm hf}$= $^{75}\!H_{\rm int}$/$^{31}\!H_{\rm int}$=3.05 in (Ca$_4$Al$_2$O$_6$)Fe$_2$(As,P)$_2$~\cite{Kinouchi_PRB_2013}. 
As shown in Fig. \ref{NMR_spectra}(b), $^{31}\!H_{\rm int}$($T$) that is proportional to $M_{\rm AFM}$ develops upon cooling below $T_{\rm N}$=35 K according to a mean-field type of $T$ dependence. The onset of AFM order is also corroborated by the peak in $(1/T_1T)$ at $T_{\rm N}$=35 K. 
Note that $M_{\rm AFM}\sim0.18\mu_{\rm B}$ and $T_{\rm N}$=35 K for the $x$=0.6 compound are smaller than $M_{\rm AFM}$=0.63~\cite{Qureshi}$\sim$0.8$\mu_{\rm B}$~\cite{H.-F.Li}  and much lower than $T_{\rm N}$=140 K for the AFM compound LaFeAsO($x$=0), respectively. 
%This may be because the shorter Fe-(As,P) bonding length for the former makes its bandwidth wider and hence the on-site electron correlation smaller than those for the latter. 

\begin{figure}[tbp]
\centering
\includegraphics[width=6.5cm]{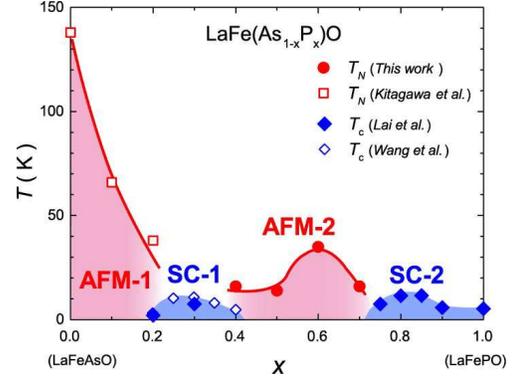}
\caption[]{(Color online) 
Phase diagram of LaFe(As$_{1-x}$P$_x$)O against $x$ in which the re-emergent AFM-2 intervenes between the SC-1 at $x <$ 0.4 (Wang {\it et al.} ~\cite{Wang}) and the SC-2 at 0.7$< x$(Lai {\it et al.}~\cite{Lai_submitted}).
} 
\label{LaFeAsPO_phase}
\end{figure}

\begin{figure}[htbp]
\centering
\includegraphics[width=6.5cm]{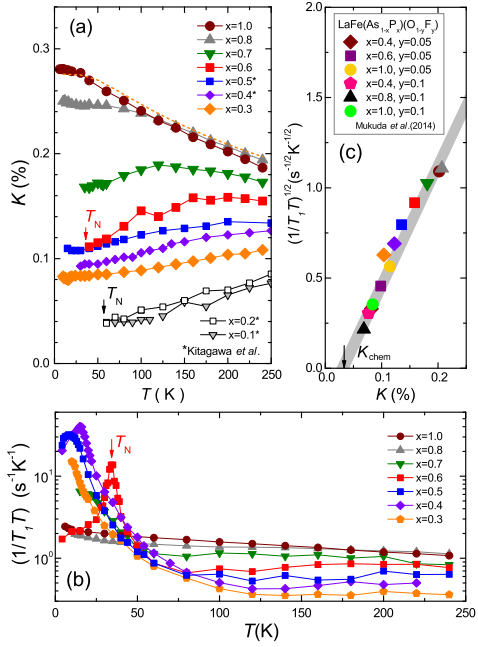}
\caption[]{(Color online) 
$T$ dependences of (a) Knight shift ($K$) and (b) $(1/T_1T)$ for LaFe(As$_{1-x}$P$_x$)O.  (c) Plot of $(1/T_1T)^{1/2}$ vs $K$ at $T$=200 K. The thick line shows the linear relation $(1/T_1T)^{1/2}=K_{\rm chem}+K_s(T)$ with $K_{\rm chem}\sim$0.03($\pm$0.01)\%.  
%$K_{\rm chem}$ was evaluated to be 0.03($\pm$0.01)\% for these compounds.
Some Knight shift data in (a)  are cited from Ref. 18.
%\cite{SKitagawa_2014} 
} 
\label{K_T1}
\end{figure}

Figure~\ref{NMR_spectra}(c) shows the $x$ dependence of the $^{31}$P-NMR spectrum at low temperatures. 
The $^{31}$P-NMR spectra for $x$=0.4, 0.5, and 0.7 except $x$=0.6 are not rectangular even at 1.9 K, pointing to an inevitable distribution of $M_{\rm AFM}$. 
Concomitantly, the peak in $(1/T_1T)$ at approximately $T_N$ is broader for these compounds than for the $x$=0.6 compound [see Fig. \ref{K_T1}(b)].
These results are indicative of some homogeneity of $T_{\rm N}$ in association with an inevitable distribution of P content in the sample as noted in a previous report for $x$=0.5~\cite{SKitagawa_2014}. 
In this context, the spatially averaged $T_{N}$ is tentatively evaluated as the temperature below which the full-width at half maximum (FWHM) of each spectrum rapidly increases as shown in Fig. \ref{NMR_spectra}(d).  
On the other hand, the spectra for $x$=0.3 and 0.8 do not undergo such significant broadening even at low temperatures, providing evidence that both are in a paramagnetic state. 
%whereas the AFM order takes place in $0.4\le x \le 0.7$.
The unexpected onset of AFM-2 at $0.4\le x \le 0.7$ is discontinuous from the AFM-1 emerging at $x\le 0.2$~\cite{SKitagawa_2014}.  
As a result, a novel phase diagram of LaFe(As$_{1-x}$P$_x$)O is summarized in Fig. \ref{LaFeAsPO_phase}, in which the re-emergent AFM-2 intervenes between the SC-1 at 0.2$<x \le$ 0.4~\cite{Wang} and the SC-2 at 0.7$< x$~\cite{Lai_submitted}.

%fig2------------------------------------------------
%------------------------------------------------

\begin{figure}[tbp]
\centering
\includegraphics[width=8cm]{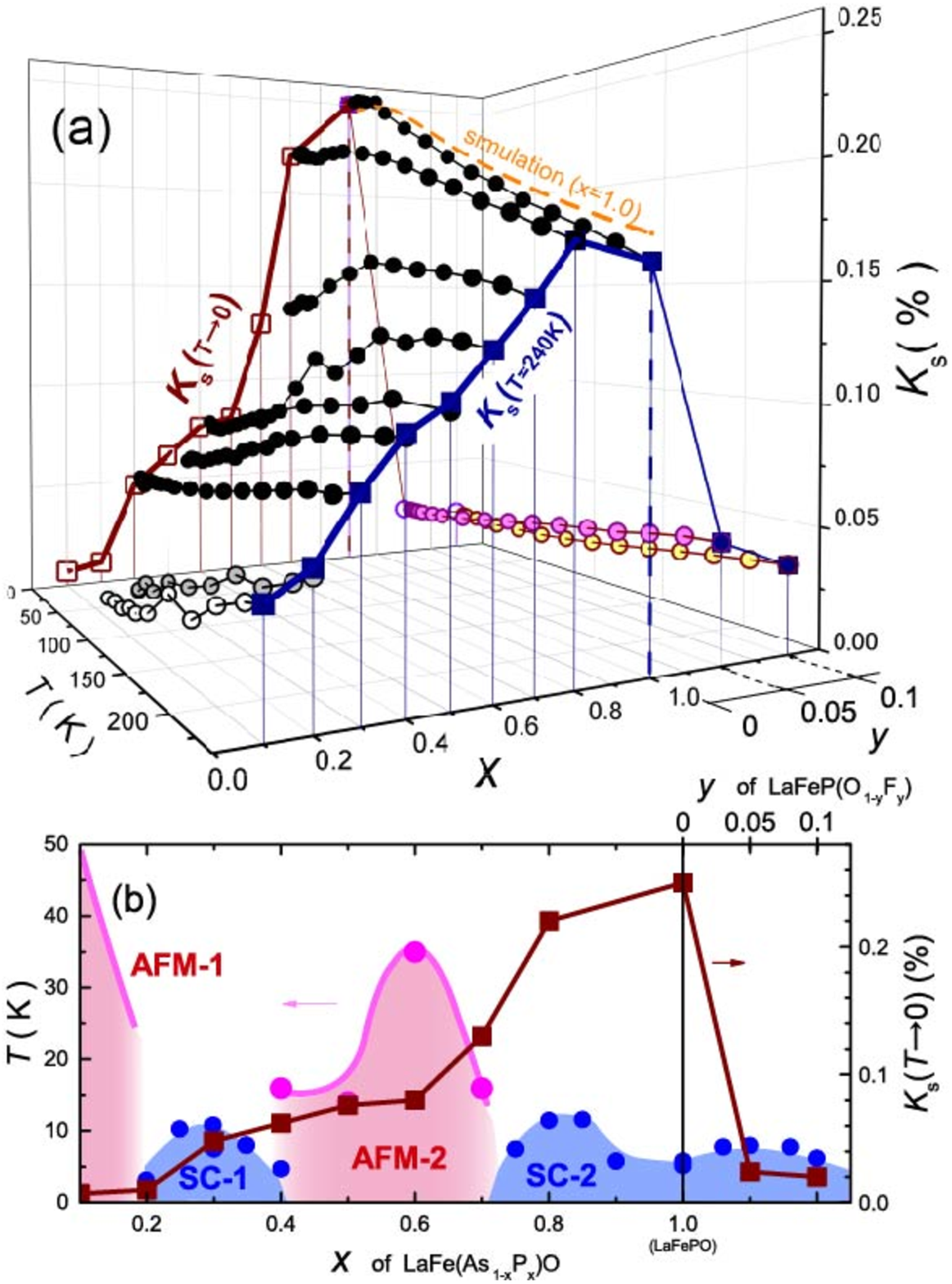}
\caption[]{(Color online) 
(a) Systematic $T$ dependences of $K_s$($T$) for each $x$ for LaFe(As$_{1-x}$P$_{x}$)O and LaFeP(O$_{1-y}$F$_y$)~\cite{Mukuda_PRB2014}, along with the $x$ dependences of $K_s$($T$=240 K) and $K_s$($T\rightarrow$0) estimated by extrapolation to $T$=0 in the paramagnetic state above $T_{\rm N}$ or $T_{\rm c}$. 
%All solid lines are eye-guides.
(b) Plot of $K_s$($T\rightarrow$0) in the phase diagram~\cite{Wang,Lai,Okuda}. 
Note that $K_s$($T\rightarrow$0) increases markedly at $x$= 0.7 $\sim$ 0.8, but electron doping through F$^-$ substitution in LaFeP(O$_{1-y}$F$_{y}$) causes $K_s$($T\rightarrow$0) to decrease markedly.~\cite{Mukuda_PRB2014}
The appearance of a sharp peak of DOS is due to the $d_{3Z^2-r^2}$ orbit for  0.7$\sim$0.8 $<x$.
} 
\label{Ks}
\end{figure}

Here, we focus on a possible P-derived evolution of the electronic state in LaFe(As$_{1-x}$P$_x$)O. 
The respective $T$ dependences of the Knight shift $K$ and $(1/T_1T)$ are shown in Figs. \ref{K_T1}(a) and \ref{K_T1}(b), respectively. 
The Knight shift comprises the $T$-dependent spin shift $K_s(T)$ and the $T$-independent chemical shift $K_{\rm chem}$.  $K_s(T)$ is given by $K_s(T)\propto~^{31}\!A_{\rm hf}\chi_0\propto~^{31}\!A_{\rm hf}N(E_{\rm F})$, using the static spin susceptibility $\chi_0$ and the density of states (DOS) $N(E_{\rm F})$ at the Fermi level $E_{\rm F}$.  
In nonmagnetic compounds, it is anticipated that $K_s(T)$ is proportional to $(1/T_1T)^{1/2}$, since Korringa's relation $(1/T_1T)\propto N(E_{\rm F})^2$ holds.  
The plot of $(1/T_1T)^{1/2}$ vs $K$ at $T$=200 K in Fig.~\ref{K_T1}(c) is close to the linear relation  $(1/T_1T)^{1/2}=K_{\rm chem}+K_s(T)$ with $K_{\rm chem}\sim$0.03 ($\pm$0.01)\%. 
Figure \ref{Ks}(a) shows $K_s(T)$ ( = $K - K_{\rm chem}$) for each $x$.  
For compounds at $x$ lower than 0.4, $K_s$($T$) decreases upon cooling as in LaFeAs(O,F)  compounds~\cite{Grafe,Terasaki,Mukuda_PRB2014}. 
This is because the Fermi level is on the tail of the large peak of the DOS beneath $E_{\rm F}$~\cite{Ikeda,Miyake}. 
However, in the intermediate $x$ range of $0.4 \le x \le 0.7$, once $K_s$($T$) increases in the high-temperature range, it then decreases toward $T$=0. 
At $x$ higher than 0.8, $K_s$($T$) monotonically increases upon cooling, suggesting the appearance of a sharp peak of DOS just at $E_{\rm F}$. 

%fig3------------------------------------------------
%------------------------------------------------

The band calculation has revealed that the Fermi surface (FS) of LaFeAsO is composed of two hole FSs at $\Gamma$(0,0),  one small hole FS at $\Gamma'$($\pi$,$\pi$), and two electron FSs at $M$[($\pi$,0),(0,$\pi$)] in the unfolded FS regime; these nearly cylindrical FSs can be connected by the interband  nesting vector ${Q}$~\cite{Vildosola,Kuroki2}. 
When the pnictogen height is as small as that in LaFePO, the $\Gamma'$ mainly originating from the $d_{X^2-Y^2}$ orbit sinks below $E_{\rm F}$, although the two nearly cylindrical hole FSs at $\Gamma$ and the two electron FSs at $M$ are maintained. 
On the other hand, a three-dimensional hole FS around $Z$($\pi$,$\pi$,$\pi$) arises from the $d_{3Z^2-r^2}$ orbit, which brings about a sharp peak in the DOS just at $E_{\rm F}$~\cite{Lebegue,Miyake}. 
As shown in the broken curves in Figs. \ref{K_T1}(a) and \ref{Ks}(a), the $T$ dependence of $K_s$($T$) at $x$=1.0 can be reproduced by assuming the DOS calculated using the band structure of LaFePO~\cite{Miyake}. 
To elucidate the appearance of such a characteristic band structure of LaFePO, we deal with $K_s$($T\rightarrow$0) estimated by extrapolation to $T$=0 in the paramagnetic state above $T_{\rm N}$ or $T_{\rm c}$. 
As shown in Figs. \ref{Ks}(a) and \ref{Ks}(b), $K_s$($T\rightarrow$0) increases markedly at $x >$ 0.7$\sim$0.8, regardless of the monotonic variation in the lattice parameters~\cite{Lai_submitted}. 
Accordingly, we remark that the electronic structure of LaFe(As$_{1-x}$P$_x$)O markedly evolves at approximately  $x$=0.7$\sim$0.8.  
Note that this evolution takes place very close to the $x$ where AFM-2 disappears. 
This result suggests that the appearance of a $d_{3Z^2-r^2}$-derived three-dimensional hole pocket causes AFM-2 to be unfavorable, in association with the possible collapse of the nearly two-dimensional LaFeAsO-like band configuration. 

%fig4------------------------------------------------
%------------------------------------------------

\begin{figure}[tbp]
\centering
\includegraphics[width=6.5cm]{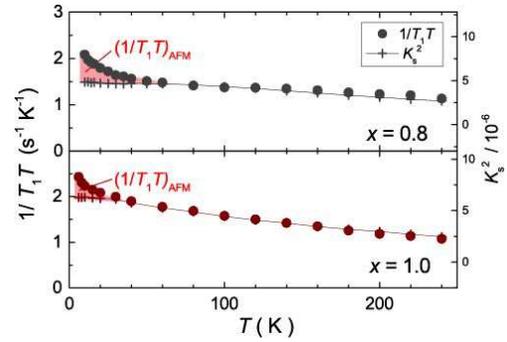}
\caption[]{(Color online) 
Evaluation of AFMSFs for $x$=0.8 and 1.0 in SC-2. 
The $1/T_1T$ of these compounds increases upon cooling below 50 K even though $K_s^2$ becomes nearly $T$-independent; hence the hatched area denoted as ($1/T_1T)_{\rm AFM}$ corresponds to the component of $1/T_1T$ due to interband scattering process, i. e., AFMSFs at a finite $Q$. 
%These results demonstrate tha is present in the minority bands. 
%To evaluate the contribution of AFM spin fluctuations, here we assumed the constant term of $1/T_1T\sim0.28$, the origin of which is unclear so far.
} 
\label{AFM-T1}
\end{figure}

Furthermore, it is noteworthy that electron doping through F$^-$ substitution for O$^{2-}$ in LaFeP(O$_{1-y}$F$_{y}$) causes $K_s$($T\rightarrow$0) to decrease markedly~\cite{Mukuda_PRB2014}, as shown in Figs. \ref{Ks}(a) and \ref{Ks}(b). 
This means that doping electrons into LaFePO forces the large DOS at $E_{\rm F}$ to shift. 
When noting that the $T_{\rm c}$ of LaFeP(O$_{1-y}$F$_{y}$) does not vary much, even though this large DOS originating from the $d_{3Z^2-r^2}$ orbit disappears with increasing $y$, the quasiparticles derived from the hole FS at the $Z$ point have nothing to do with the onset of SC-2. 
On the other hand, as shown in  Fig. \ref{AFM-T1}, $1/T_1T$ at $x$=0.8 and 1.0 increases upon cooling below 50 K even though $K_s^2$ becomes nearly $T$-independent in this $T$ range,  demonstrating that the subtle evolution of AFMSFs at a finite ${Q}$ is present  in other minority bands in SC-2. 
Taking the result of the band calculation into account, we suggest that the quasiparticles originating from the $d_{XZ}$/$d_{YZ}$-derived nearly two-dimensional minority bands are responsible for the SC-2 with a relatively low  $T_{\rm c}$. 
In this context, this tendency between the onset of  $T_{\rm c}$ and the diversity of the band structure is indicative of the multiband nature observed in Fe-pnictide SC in general.

%fig5------------------------------------------------
%------------------------------------------------

Finally, we deal with some differences of AFM-1 and AFM-2 in LaFe(As$_{1-x}$P$_x$)O.
The AFM-1 of LaFeAsO at $x$=0 disappears rapidly around $x\sim$0.2 where $h_{\rm Pn}$ is approximately 1.29 \AA, which coincides with that at the border between the AFM and SC phases observed in many Fe pnictides in the Fe$^{2+}$-like state with neither electron nor hole doping~\cite{Kinouchi_PRB_2013}.  
AFM-2 appears in the $h_{\rm Pn}$ range of 1.2$\sim$1.25 \AA, which is smaller than 1.29 \AA.  
From the recent band calculation, Kuroki {\it et al.} suggested that FS nesting at bands mainly composed of  $d_{XZ}$/$d_{YZ}$ orbits becomes better again at the intermediate $x$ of LaFe(As$_{1-x}$P$_x$)O~\cite{Kuroki_private_com},  consistent with the experimental results. 
From other context, the reason why the compound LaFe(As$_{1-x}$P$_x$)(O$_{0.95}$F$_{0.05}$) with $x$=0.6 exhibits a maximum  $T_{\rm c}$=24 K against $x$ is found, that is,  the marked enhancement of  AFMSFs at low energies as a result of the depression of AFM-2 by electron doping~\cite{Mukuda_PRB2014}. 
Eventually, interband nesting on $d_{XZ}$/$d_{YZ}$ orbits must be a key factor for the emergence of SC around AFM-2.  

However, note that AFMSFs at low energies are not always highly significant in Fe pnictide compounds with $T_{\rm c}>$ 50 K~\cite{MukudaPRL}. 
Recently, another type of stripe AFM(H) phase carrying a large AFM moment but exhibiting a low $T_{\rm N}$  has been reported for heavily electron-doped compounds LaFeAs(O,H)~\cite{Hiraishi}: These electronic states undergoing a novel structural deformation are totally different from those of the mother compound with AFM-1 order~\cite{Iimura_H}. 
It has been reported that the SC(H) in LaFeAs(O,H) that exhibits a $T_{\rm c}$ higher than in SC-1 occurs owing to the development of AFMSFs at high energies in the vicinity of the AFM(H) phase despite the FS nesting condition being significantly worse~\cite{Suzuki_50K}. 
In this point, the origins of the AFM and SC phases are yet unresolved underlying issues in LaFeAsO-based compounds, which should be clarified through a unified picture in the near future.

In summary,  the present $^{31}$P-NMR studies of LaFe(As$_{1-x}$P$_x$)O  have unraveled the re-emergence of AFM-2 with a homogeneous moment of $M_{\rm AFM}\sim$0.18$\mu_{\rm B}$ and $T_{\rm N}$=35 K at $x$=0.6.  
It is highlighted that this AFM-2 takes place at 0.4$\le x \le$0.7, intervening between the SC-1 at 0.2$<x<$ 0.4 and the SC-2 at 0.7$<x$. 
Note that the electronic structure of LaFe(As$_{1-x}$P$_x$)O markedly evolves at $x$=0.7$\sim$0.8 in such a manner that a three-dimensional hole FS around $Z$($\pi$, $\pi$, $\pi$) arising from the $d_{3Z^2-r^2}$ orbit brings about a sharp peak in the DOS just at $E_{\rm F}$~\cite{Lebegue,Miyake}. 
This result suggests that the appearance of a $d_{3Z^2-r^2}$-derived three-dimensional hole pocket causes AFM-2 to be unfavorable in association with the collapse of the nearly two-dimensional LaFeAsO-like band configuration. 
From another context, the reason why the compound LaFe(As$_{1-x}$P$_x$)(O$_{0.95}$F$_{0.05}$) with $x$=0.6 exhibits a maximum  $T_{\rm c}$=24 K against $x$ has been found, that is, the marked enhancement of AFMSFs at low energies as a result of the depression of AFM-2 by electron doping~\cite{Mukuda_PRB2014}. 
Eventually, interband nesting on $d_{XZ}$/$d_{YZ}$ orbits must be a key factor for the emergence of SC around AFM-2.

%\section*{Acknowledgements}

{\footnotesize We thank K. Kuroki, H. Usui, and K. Suzuki for valuable discussion, and K. Ishida  for providing us their experimental data in the low $x$ range. This work was supported by KAKENHI from JSPS.}

%::::::::::::::::bibliography::::::::::::::::::::::::::::::::::::::::::::::::
%:::::::::::::::::::::::::::::::::::::::::::::::::::::::

\end{document}